\documentclass[12pt]{article}
\usepackage{emlines2}
\usepackage{graphicx}
\newcommand{\m}{\mbox{\boldmath $m$}}

\newcommand{\n}{\mbox{\boldmath $n$}}
\newcommand{\sn}{\mbox{\scriptsize\boldmath $n$}}

\newcommand{\sta}{\mbox{\scriptsize\boldmath $\tau$}}
\newcommand{\ta}{\mbox{\boldmath $\tau$}}

\newcommand{\rr}{\mbox{\boldmath $r$}}

\newcommand{\p}{\mbox{\boldmath $p$}}

\newcommand{\kk}{\mbox{\boldmath $k$}}
\begin{document}
\begin{center}
{\bf Dynamical diffraction}
\medskip

V.K.Ignatovich
\medskip

FLNP, Joint Institute for Nuclear Research,  141980, Dubna, Russia.
\medskip

\end{center}
\begin{abstract}
An algebraic approach to neutron scattering on a one dimensional potentials is generalized to
diffraction on three dimensional single crystals.
\end{abstract}
\begin{flushright}
{\small I hate and will always hate dynamical diffraction\\
R.Golub}
\end{flushright}

\section{Introduction}
I feel a solidarity with R.Golub with respect to the standard dynamical diffraction theory,
because for me it was always so boring to explain diffraction effects by motion of some points
along mathematical dispersion
surfaces. Here I want to present an alternative theory, which has a clear physical meaning at every step.
My approach is similar to that of~\cite{dar,god}.

\subsection{Algebra of neutron scattering on two semitransparent mirrors}

\begin{figure}[b]
\hspace{20mm}
\special{em:linewidth 0.4pt}
\unitlength 0.60mm
\linethickness{0.4pt}
\begin{picture}(202.67,47.00)
\emline{2.66}{11.00}{1}{139.00}{11.00}{2}
\emline{22.33}{11.00}{3}{23.52}{13.79}{4}
\emline{23.52}{13.79}{5}{24.70}{16.35}{6}
\emline{24.70}{16.35}{7}{25.87}{18.67}{8}
\emline{25.87}{18.67}{9}{27.04}{20.76}{10}
\emline{27.04}{20.76}{11}{28.21}{22.61}{12}
\emline{28.21}{22.61}{13}{29.37}{24.23}{14}
\emline{29.37}{24.23}{15}{30.53}{25.61}{16}
\emline{30.53}{25.61}{17}{31.68}{26.75}{18}
\emline{31.68}{26.75}{19}{32.83}{27.66}{20}
\emline{32.83}{27.66}{21}{33.97}{28.34}{22}
\emline{33.97}{28.34}{23}{35.11}{28.78}{24}
\emline{35.11}{28.78}{25}{36.25}{28.98}{26}
\emline{36.25}{28.98}{27}{37.38}{28.95}{28}
\emline{37.38}{28.95}{29}{38.51}{28.69}{30}
\emline{38.51}{28.69}{31}{39.63}{28.19}{32}
\emline{39.63}{28.19}{33}{40.75}{27.45}{34}
\emline{40.75}{27.45}{35}{41.86}{26.48}{36}
\emline{41.86}{26.48}{37}{42.97}{25.27}{38}
\emline{42.97}{25.27}{39}{44.08}{23.83}{40}
\emline{44.08}{23.83}{41}{45.18}{22.15}{42}
\emline{45.18}{22.15}{43}{46.27}{20.24}{44}
\emline{46.27}{20.24}{45}{47.37}{18.09}{46}
\emline{47.37}{18.09}{47}{48.45}{15.71}{48}
\emline{48.45}{15.71}{49}{50.33}{11.00}{50}
\emline{144.66}{11.00}{51}{146.12}{13.34}{52}
\emline{146.12}{13.34}{53}{147.59}{15.47}{54}
\emline{147.59}{15.47}{55}{149.06}{17.38}{56}
\emline{149.06}{17.38}{57}{150.54}{19.08}{58}
\emline{150.54}{19.08}{59}{152.02}{20.56}{60}
\emline{152.02}{20.56}{61}{153.52}{21.83}{62}
\emline{153.52}{21.83}{63}{155.01}{22.89}{64}
\emline{155.01}{22.89}{65}{156.52}{23.73}{66}
\emline{156.52}{23.73}{67}{158.03}{24.36}{68}
\emline{158.03}{24.36}{69}{159.54}{24.78}{70}
\emline{159.54}{24.78}{71}{161.07}{24.98}{72}
\emline{161.07}{24.98}{73}{162.60}{24.97}{74}
\emline{162.60}{24.97}{75}{164.13}{24.74}{76}
\emline{164.13}{24.74}{77}{165.67}{24.30}{78}
\emline{165.67}{24.30}{79}{167.22}{23.65}{80}
\emline{167.22}{23.65}{81}{168.78}{22.78}{82}
\emline{168.78}{22.78}{83}{170.34}{21.70}{84}
\emline{170.34}{21.70}{85}{171.91}{20.41}{86}
\emline{171.91}{20.41}{87}{173.48}{18.90}{88}
\emline{173.48}{18.90}{89}{175.06}{17.18}{90}
\emline{175.06}{17.18}{91}{176.65}{15.25}{92}
\emline{176.65}{15.25}{93}{178.24}{13.10}{94}
\emline{178.24}{13.10}{95}{179.66}{11.00}{96}
\put(35.33,16.67){\makebox(0,0)[cc]{$V_1$}}
\put(163.00,16.33){\makebox(0,0)[cc]{$V_2$}}
\put(23.00,15.33){\vector(1,0){0.2}}
\emline{-3.00}{15.33}{97}{23.00}{15.33}{98}
\put(11.00,19.67){\makebox(0,0)[cc]{$1$}}
\put(193.33,5.67){\makebox(0,0)[cc]{$x$}}
\put(105.33,5.33){\makebox(0,0)[cc]{$l$}}
\put(145.00,5.33){\vector(1,0){0.2}}
\emline{134.33}{5.33}{99}{145.00}{5.33}{100}
\put(50.00,5.33){\vector(-1,0){0.2}}
\emline{63.33}{5.33}{101}{50.00}{5.33}{102}
\put(-1.67,28.00){\vector(-1,0){0.2}}
\emline{23.66}{28.00}{103}{-1.67}{28.00}{104}
\put(11.00,34.67){\makebox(0,0)[cc]{$R_{12}$}}
\put(202.66,18.67){\vector(1,0){0.2}}
\emline{182.00}{18.67}{105}{202.66}{18.67}{106}
\put(192.33,23.67){\makebox(0,0)[cc]{$T_{12}$}}
\put(22.66,5.00){\makebox(0,0)[cc]{0}}
\put(179.33,5.00){\makebox(0,0)[cc]{$s$}}
\put(145.00,15.33){\vector(1,0){0.2}}
\emline{134.00}{15.33}{107}{145.00}{15.33}{108}
\put(133.00,28.00){\vector(-1,0){0.2}}
\emline{145.66}{28.00}{109}{133.00}{28.00}{110}
\put(50.33,28.00){\vector(-1,0){0.2}}
\emline{60.33}{28.00}{111}{50.33}{28.00}{112}
\put(141.00,19.67){\makebox(0,0)[cc]{$X$}}
\put(139.33,34.67){\makebox(0,0)[cc]{$R_2X$}}
\put(54.00,34.67){\makebox(0,0)[cc]{$e^{ikl}R_2X$}}
\put(53.00,19.67){\makebox(0,0)[lc]{$T_1+R_1e^{ikl}R_2X$}}
\emline{115.33}{5.33}{113}{136.00}{5.33}{114}
\emline{64.00}{5.33}{115}{98.66}{5.33}{116}
\emline{139.33}{10.67}{117}{202.67}{10.67}{118}
\put(86.67,15.33){\vector(1,0){0.2}}
\emline{50.67}{15.33}{119}{86.67}{15.33}{120}
\end{picture}
\caption{\label{2barx} Two semitransparent mirrors are represented by
two potentials separated by distance $l$.}
\end{figure}
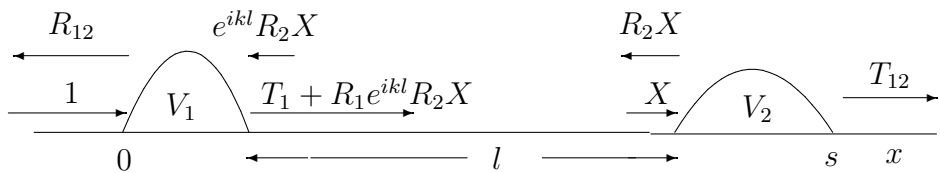

To explain the approach, let's consider reflection and transmission of a
system of two separated semitransparent mirrors.
We consider our primary neutron to be described by a plane wave
$\psi_0(\kk,\rr,t)$, and two mirrors --- by two different potentials
$u_{1,2}$, separated by a distance $l$ as shown in fig. \ref{2barx}.
If
we know reflection $r_{1,2}$ and transmission $t_{1,2}$ amplitudes
of every separate mirror, we can find reflection $R_{12}$
and transmission $T_{12}$ of both mirrors.

Suppose the amplitude of the incident wave $\exp(ikx)$ on the surface of the
first mirror at $x=0$ is equal to unity at the surface $x=0$ of the
first mirror.
Let's denote $X$ the amplitude of the wave incident on the surface of the second mirror
(see fig. \ref{2barx}).
For $X$ we can construct an equation
\begin{equation}
X=\exp(ikl)t_1+\exp(ikl)r_1\exp(ikl)r_2X.
\label{ee2}
\end{equation}
The second wave in the right hand side is generated by the $X$:
the incident wave with amplitude $X$ reflects from the second mirror,
goes to the first one, reflects from the first mirror,
and arrives to the second one. During propagations between mirrors the wave acquires the phase factor
$\exp(ikl)$. At the second mirror the second term of (\ref{ee2}) superposes
with the wave, which is transmitted through
the first mirror and is represented by the first term in the right hand side.
The sum of two waves according to definition is equal to $X$.
\bigskip

Solution of the Eq. (\ref{ee2}) is
\begin{equation}
X=\frac{t_1\exp(ikl)}{1-r_1r_2\exp(2ikl)}.
\label{ee3}
\end{equation}
Reflected and transmitted waves can now be deduced as
\begin{equation}
R_{12}=r_1+t_1\exp(ikl)r_2X,\qquad T_{12}=t_2X.
\label{ee4}
\end{equation}
Reflected wave is a superposition of two waves. The first one is related to the wave directly reflected
from the first mirror,
and the second one, is related to the wave $X$, which after reflection from the second mirror arrives to the first
one and transmits through it.
Substitution of $X$ into (\ref{ee3}) gives
\begin{equation}
R_{12}=r_1+t_1^2\frac{e^{2ikl}r_2}{1-e^{2ikl}r_2r_1},\qquad
T_{12}=t_1t_2\frac{e^{ikl}}{1-e^{2ikl}r_2r_1}.
\label{r12}
\end{equation}

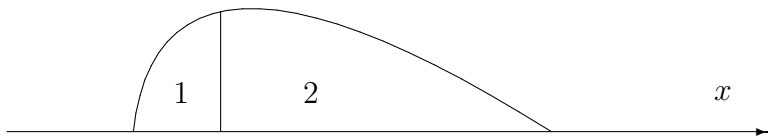
\begin{figure}[bh]
\vspace{-1.5cm}

\hspace{10mm}
\special{em:linewidth 0.4pt}
\unitlength 1.20mm
\linethickness{0.4pt}
\begin{picture}(103.00,28.34)
\put(103.00,1.00){\vector(1,0){0.2}}
\emline{19.00}{1.00}{1}{103.00}{1.00}{2}
\emline{33.00}{1.00}{3}{33.30}{3.09}{4}
\emline{33.30}{3.09}{5}{33.74}{5.01}{6}
\emline{33.74}{5.01}{7}{34.30}{6.76}{8}
\emline{34.30}{6.76}{9}{34.99}{8.33}{10}
\emline{34.99}{8.33}{11}{35.80}{9.73}{12}
\emline{35.80}{9.73}{13}{36.75}{10.95}{14}
\emline{36.75}{10.95}{15}{37.82}{12.00}{16}
\emline{37.82}{12.00}{17}{39.03}{12.88}{18}
\emline{39.03}{12.88}{19}{40.36}{13.58}{20}
\emline{40.36}{13.58}{21}{41.82}{14.11}{22}
\emline{41.82}{14.11}{23}{43.40}{14.46}{24}
\emline{43.40}{14.46}{25}{45.12}{14.64}{26}
\emline{45.12}{14.64}{27}{46.97}{14.65}{28}
\emline{46.97}{14.65}{29}{48.94}{14.48}{30}
\emline{48.94}{14.48}{31}{51.04}{14.14}{32}
\emline{51.04}{14.14}{33}{53.27}{13.62}{34}
\emline{53.27}{13.62}{35}{55.63}{12.93}{36}
\emline{55.63}{12.93}{37}{58.12}{12.06}{38}
\emline{58.12}{12.06}{39}{60.73}{11.02}{40}
\emline{60.73}{11.02}{41}{63.48}{9.81}{42}
\emline{63.48}{9.81}{43}{66.35}{8.42}{44}
\emline{66.35}{8.42}{45}{69.35}{6.86}{46}
\emline{69.35}{6.86}{47}{72.48}{5.12}{48}
\emline{72.48}{5.12}{49}{75.74}{3.21}{50}
\emline{75.74}{3.21}{51}{79.33}{1.00}{52}
\emline{42.67}{14.34}{53}{42.67}{1.00}{54}
\put(38.33,5.34){\makebox(0,0)[cc]{1}}
\put(52.67,5.34){\makebox(0,0)[cc]{2}}
\put(98.33,5.34){\makebox(0,0)[cc]{$x$}}
\end{picture}
\caption{\label{nonsym} Every potential can be split into two parts at an arbitrary
point.}
\end{figure}

Two mirrors above were separated by distance $l$. It is evident that (\ref{r12}) are
valid for arbitrary $l$. In particular $l$ can be zero, and with such $l$ we have
\begin{equation}
R_{12}=r_1+t_1^2\frac{r_2}{1-r_1r_2},\qquad T_{12}=\frac{t_2t_1}{1-r_1r_2}.
\label{tr}
\end{equation}
This result is a discovery~\cite{ig0}. It shows that an arbitrary potential can at any point be split
into two parts.
If both parts are symmetrical, we get (\ref{tr}).
However the potential and its parts, as is shown in fig. \ref{nonsym}, are in general nonsymmetrical.
In that case reflection $r_i$ and transmission $t_i$ amplitudes from the left and from the right
are in general different, so formulas (\ref{tr}) are to be changed as follows
\begin{equation}
\overrightarrow R_{12}=\overrightarrow{r_1}+
\overleftarrow{t_1}\frac{\overrightarrow{r_2}}
{1-\overleftarrow{r_1}\overrightarrow{r_2}}\overrightarrow{t_1},\qquad
\overleftarrow R_{21}=\overleftarrow{r_2}+
\overrightarrow{t_2}\frac{\overleftarrow{r_1}}
{1-\overleftarrow{r_1}\overrightarrow{r_2}}\overleftarrow{t_2},
\label{xx}
\end{equation}
and
\begin{equation}
\overrightarrow T_{12}=\frac{\overrightarrow{t_2}\overrightarrow{t_1}}
{1-\overleftarrow{r_1}\overrightarrow{r_2}},\qquad
\overleftarrow T_{21}=\frac{\overleftarrow{t_1}\overleftarrow{t_2}}
{1-\overleftarrow{r_1}\overrightarrow{r_2}},
\label{xx2}
\end{equation}
where arrow shows the direction of the incident wave.

\subsection{Properties of reflection and transmission amplitudes}

\subsubsection{Nonsymmetric potentials}

\begin{figure}[t]
\special{em:linewidth 0.4pt}
\unitlength 0.80mm
\linethickness{0.4pt}
\hspace{3cm}\begin{picture}(117.33,47.00)
\emline{29.00}{11.00}{1}{30.18}{13.79}{2}
\emline{30.18}{13.79}{3}{31.36}{16.35}{4}
\emline{31.36}{16.35}{5}{32.54}{18.67}{6}
\emline{32.54}{18.67}{7}{33.71}{20.76}{8}
\emline{33.71}{20.76}{9}{34.88}{22.61}{10}
\emline{34.88}{22.61}{11}{36.04}{24.23}{12}
\emline{36.04}{24.23}{13}{37.19}{25.61}{14}
\emline{37.19}{25.61}{15}{38.35}{26.75}{16}
\emline{38.35}{26.75}{17}{39.50}{27.66}{18}
\emline{39.50}{27.66}{19}{40.64}{28.34}{20}
\emline{40.64}{28.34}{21}{41.78}{28.78}{22}
\emline{41.78}{28.78}{23}{42.92}{28.98}{24}
\emline{42.92}{28.98}{25}{44.05}{28.95}{26}
\emline{44.05}{28.95}{27}{45.17}{28.69}{28}
\emline{45.17}{28.69}{29}{46.30}{28.19}{30}
\emline{46.30}{28.19}{31}{47.42}{27.45}{32}
\emline{47.42}{27.45}{33}{48.53}{26.48}{34}
\emline{48.53}{26.48}{35}{49.64}{25.27}{36}
\emline{49.64}{25.27}{37}{50.74}{23.83}{38}
\emline{50.74}{23.83}{39}{51.84}{22.15}{40}
\emline{51.84}{22.15}{41}{52.94}{20.24}{42}
\emline{52.94}{20.24}{43}{54.03}{18.09}{44}
\emline{54.03}{18.09}{45}{55.12}{15.71}{46}
\emline{55.12}{15.71}{47}{57.00}{11.00}{48}
\emline{56.66}{11.00}{49}{58.12}{13.34}{50}
\emline{58.12}{13.34}{51}{59.59}{15.47}{52}
\emline{59.59}{15.47}{53}{61.06}{17.38}{54}
\emline{61.06}{17.38}{55}{62.54}{19.08}{56}
\emline{62.54}{19.08}{57}{64.02}{20.56}{58}
\emline{64.02}{20.56}{59}{65.52}{21.83}{60}
\emline{65.52}{21.83}{61}{67.01}{22.89}{62}
\emline{67.01}{22.89}{63}{68.52}{23.73}{64}
\emline{68.52}{23.73}{65}{70.03}{24.36}{66}
\emline{70.03}{24.36}{67}{71.54}{24.78}{68}
\emline{71.54}{24.78}{69}{73.07}{24.98}{70}
\emline{73.07}{24.98}{71}{74.60}{24.97}{72}
\emline{74.60}{24.97}{73}{76.13}{24.74}{74}
\emline{76.13}{24.74}{75}{77.67}{24.30}{76}
\emline{77.67}{24.30}{77}{79.22}{23.65}{78}
\emline{79.22}{23.65}{79}{80.78}{22.78}{80}
\emline{80.78}{22.78}{81}{82.34}{21.70}{82}
\emline{82.34}{21.70}{83}{83.91}{20.41}{84}
\emline{83.91}{20.41}{85}{85.48}{18.90}{86}
\emline{85.48}{18.90}{87}{87.06}{17.18}{88}
\emline{87.06}{17.18}{89}{88.65}{15.25}{90}
\emline{88.65}{15.25}{91}{90.24}{13.10}{92}
\emline{90.24}{13.10}{93}{91.66}{11.00}{94}
\put(42.00,16.67){\makebox(0,0)[cc]{$u_1$}}
\put(75.00,16.33){\makebox(0,0)[cc]{$u_2$}}
\put(29.67,15.33){\vector(1,0){0.2}}
\emline{3.67}{15.33}{95}{29.67}{15.33}{96}
\put(17.67,19.67){\makebox(0,0)[cc]{$e^{ikx}$}}
\put(105.33,5.67){\makebox(0,0)[cc]{$x$}}
\put(5.00,28.00){\vector(-1,0){0.2}}
\emline{30.33}{28.00}{97}{5.00}{28.00}{98}
\put(17.67,34.67){\makebox(0,0)[cc]{$R_{12}e^{-ikx}$}}
\put(114.66,18.67){\vector(1,0){0.2}}
\emline{94.00}{18.67}{99}{114.66}{18.67}{100}
\put(104.33,23.67){\makebox(0,0)[cc]{$T_{12}e^{ik(x-s)}$}}
\put(29.33,5.00){\makebox(0,0)[cc]{0}}
\put(91.33,5.00){\makebox(0,0)[cc]{$s$}}
\put(117.33,10.00){\vector(1,0){0.2}}
\emline{3.00}{10.00}{101}{117.33}{10.00}{102}
\end{picture}
\caption{\label{2barnsep} A nonsymmetric potential constructed with the help of two symmetrical ones.}
\end{figure}
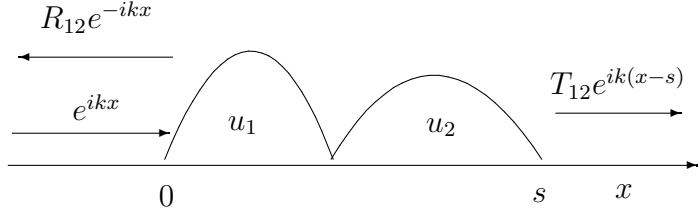

To find the difference of amplitudes for waves incident from the left and from the right it is
sufficient to consider the simplest nonsymmetric potential, which consists of two symmetrical ones
as shown in fig. \ref{2barnsep}.

Transmission and reflection of the total potential are given by
(\ref{tr}). From them we can immediately see that
$T_{12}\equiv T_{21}$, therefore for arbitrary nonsymmetrical potential we have
$\overleftarrow t=\overrightarrow t=t$.
Reflection amplitudes in general are not equal to each other, however in the case of negligible losses
they differ only by a phase factor, i.e.
$\overleftarrow r=\exp(i\chi)\overrightarrow r$,
where $\chi$ is a real number.
Indeed, because of unitarity we have
\begin{equation}
|\overleftarrow r|^2+|t|^2=|\overrightarrow r|^2+
|t|^2=1,
\label{norm}
\end{equation}
therefore
$|\overleftarrow r|=|\overrightarrow r|$.

\subsubsection{Relation between phases of $r$ and $t$ for symmetrical potentials}

The amplitudes $r$ and $t$ of a symmetrical potential
can be represented as $r=|r|\exp(i\phi_r)$ and $t=|t|\exp(i\phi_t)$. It is easy to show by a gedanken
experiment~\cite{ig}
that for real potentials
\begin{equation}\label{ph}
\phi_r=\phi_t\pm\pi/2.
\end{equation}
It follows that $t=\pm i|t|e^{i\phi_r}$, combinations $r\pm t$ are unit complex numbers,
and $r^2-t^2=e^{2i\phi_r}$.

For nonsymmetrical potentials we have
$\overrightarrow r=|r|\exp(i\overrightarrow\chi_r)$,
$\overleftarrow r=|r|\exp(i\overleftarrow\chi_r)$, and instead of
(\ref{ph}) we can obtain
$(\overrightarrow\chi_r+\overleftarrow\chi_r)/2=\chi_t\pm\pi/2$.

\subsection{An example of application of the above algebra}

\begin{figure}[b]
\special{em:linewidth 0.4pt}
\unitlength 0.80mm
\linethickness{0.4pt}
\begin{picture}(150.00,38.00)
\put(139.33,7.34){\vector(1,0){0.2}}
\emline{60.67}{7.34}{1}{139.33}{7.34}{2}
\emline{95.67}{7.34}{3}{95.67}{28.34}{4}
\put(59.00,35.00){\makebox(0,0)[cc]{$u$}}
\emline{67.67}{28.34}{5}{65.67}{28.34}{6}
\put(59.00,28.34){\makebox(0,0)[cc]{$u_0$}}
\put(135.33,2.00){\makebox(0,0)[cc]{$x$}}
\put(95.67,2.00){\makebox(0,0)[cc]{0}}
\put(93.00,11.67){\vector(1,0){0.2}}
\emline{63.33}{11.67}{7}{93.00}{11.67}{8}
\put(86.33,15.67){\makebox(0,0)[cc]{$e^{ikx}$}}
\put(65.67,37.00){\vector(0,1){0.2}}
\emline{65.67}{7.33}{9}{65.67}{37.00}{10}
\emline{95.67}{28.33}{11}{126.00}{28.33}{12}
\emline{126.00}{28.33}{13}{126.00}{7.33}{14}
\put(126.00,1.67){\makebox(0,0)[cc]{$d$}}
\put(150.00,11.67){\vector(1,0){0.2}}
\emline{131.00}{11.67}{15}{150.00}{11.67}{16}
\put(140.00,15.67){\makebox(0,0)[cc]{$te^{ik(x-d)}$}}
\put(126.00,31.67){\vector(1,0){0.2}}
\emline{113.67}{31.67}{17}{126.00}{31.67}{18}
\put(119.67,38.00){\makebox(0,0)[cc]{$X$}}
\put(63.00,23.67){\vector(-1,0){0.2}}
\emline{93.00}{23.67}{19}{63.00}{23.67}{20}
\put(78.67,27.00){\makebox(0,0)[cc]{$re^{-ikx}$}}
\end{picture}
\caption{\label{rect} Rectangular potential barrier of height $u_0$ and width
$d$}
\end{figure}
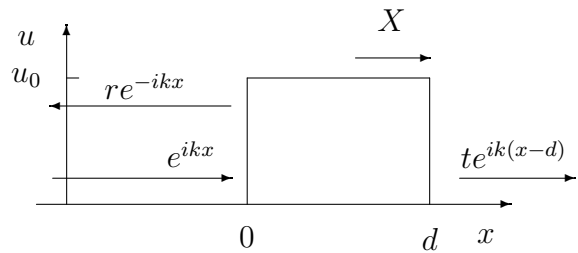

One of the simplest is a rectangular potential shown in fig. \ref{rect}.
\label{srec}
If we denote the amplitude of wave incident upon the right edge of the rectangle from inside the barrier,
we can find an equation for $X$
\begin{equation}
X=\exp(ik'd)t_0+\exp(ik'd)(-r_0)\exp(ik'd)
(-r_0)X,
\label{ste8}
\end{equation}
where
\begin{equation}
r_0=\frac{k-k'}{k+k'},\qquad t_0
=\frac{2k}{k-k'},
\label{ste9}
\end{equation}
and $k'=\sqrt{k^2-u}$ is the particle wave number inside the potential.

The eq. (\ref{ste8}) is easily solved. With this $X$ we find both reflection
$R$ and transmission $T$ amplitudes of rectangle. We omit here all the
mathematics (look, for instance, in~\cite{god,ig0})  and put down
the final result
\begin{equation}
R=r_0\frac{1-\exp(2ik'd)}
{1-r_0^2\exp(2ik'd)},\qquad T=e^{ik'd}\frac{1-r_0^2}{1-r_0^2\exp(2ik'd)},
\label{ste12}
\end{equation}
which will be used later.

We restrict here ourselves with only this example of application of above algebra,
but those, who want to see more examples, can find them in
references~\cite{ig}---~\cite{korn3}

\section{One dimensional periodic potential}

\begin{figure}[b]
\special{em:linewidth 0.4pt}
\unitlength 0.80mm
\linethickness{0.4pt}
\begin{picture}(163.34,26.33)
\emline{68.67}{5.33}{1}{70.06}{7.46}{2}
\emline{70.06}{7.46}{3}{71.40}{9.32}{4}
\emline{71.40}{9.32}{5}{72.67}{10.90}{6}
\emline{72.67}{10.90}{7}{73.87}{12.22}{8}
\emline{73.87}{12.22}{9}{75.01}{13.27}{10}
\emline{75.01}{13.27}{11}{76.09}{14.05}{12}
\emline{76.09}{14.05}{13}{77.10}{14.57}{14}
\emline{77.10}{14.57}{15}{78.05}{14.81}{16}
\emline{78.05}{14.81}{17}{78.94}{14.78}{18}
\emline{78.94}{14.78}{19}{79.76}{14.49}{20}
\emline{79.76}{14.49}{21}{80.52}{13.92}{22}
\emline{80.52}{13.92}{23}{81.22}{13.09}{24}
\emline{81.22}{13.09}{25}{81.85}{11.98}{26}
\emline{81.85}{11.98}{27}{82.42}{10.61}{28}
\emline{82.42}{10.61}{29}{82.92}{8.97}{30}
\emline{82.92}{8.97}{31}{83.67}{5.33}{32}
\emline{83.34}{5.33}{33}{84.73}{7.46}{34}
\emline{84.73}{7.46}{35}{86.07}{9.32}{36}
\emline{86.07}{9.32}{37}{87.34}{10.90}{38}
\emline{87.34}{10.90}{39}{88.54}{12.22}{40}
\emline{88.54}{12.22}{41}{89.68}{13.27}{42}
\emline{89.68}{13.27}{43}{90.76}{14.05}{44}
\emline{90.76}{14.05}{45}{91.77}{14.57}{46}
\emline{91.77}{14.57}{47}{92.72}{14.81}{48}
\emline{92.72}{14.81}{49}{93.61}{14.78}{50}
\emline{93.61}{14.78}{51}{94.43}{14.49}{52}
\emline{94.43}{14.49}{53}{95.19}{13.92}{54}
\emline{95.19}{13.92}{55}{95.89}{13.09}{56}
\emline{95.89}{13.09}{57}{96.52}{11.98}{58}
\emline{96.52}{11.98}{59}{97.09}{10.61}{60}
\emline{97.09}{10.61}{61}{97.59}{8.97}{62}
\emline{97.59}{8.97}{63}{98.34}{5.33}{64}
\emline{98.34}{5.33}{65}{99.73}{7.46}{66}
\emline{99.73}{7.46}{67}{101.07}{9.32}{68}
\emline{101.07}{9.32}{69}{102.34}{10.90}{70}
\emline{102.34}{10.90}{71}{103.54}{12.22}{72}
\emline{103.54}{12.22}{73}{104.68}{13.27}{74}
\emline{104.68}{13.27}{75}{105.76}{14.05}{76}
\emline{105.76}{14.05}{77}{106.77}{14.57}{78}
\emline{106.77}{14.57}{79}{107.72}{14.81}{80}
\emline{107.72}{14.81}{81}{108.61}{14.78}{82}
\emline{108.61}{14.78}{83}{109.43}{14.49}{84}
\emline{109.43}{14.49}{85}{110.19}{13.92}{86}
\emline{110.19}{13.92}{87}{110.89}{13.09}{88}
\emline{110.89}{13.09}{89}{111.52}{11.98}{90}
\emline{111.52}{11.98}{91}{112.09}{10.61}{92}
\emline{112.09}{10.61}{93}{112.59}{8.97}{94}
\emline{112.59}{8.97}{95}{113.34}{5.33}{96}
\emline{113.34}{5.33}{97}{114.73}{7.46}{98}
\emline{114.73}{7.46}{99}{116.07}{9.32}{100}
\emline{116.07}{9.32}{101}{117.34}{10.90}{102}
\emline{117.34}{10.90}{103}{118.54}{12.22}{104}
\emline{118.54}{12.22}{105}{119.68}{13.27}{106}
\emline{119.68}{13.27}{107}{120.76}{14.05}{108}
\emline{120.76}{14.05}{109}{121.77}{14.57}{110}
\emline{121.77}{14.57}{111}{122.72}{14.81}{112}
\emline{122.72}{14.81}{113}{123.61}{14.78}{114}
\emline{123.61}{14.78}{115}{124.43}{14.49}{116}
\emline{124.43}{14.49}{117}{125.19}{13.92}{118}
\emline{125.19}{13.92}{119}{125.89}{13.09}{120}
\emline{125.89}{13.09}{121}{126.52}{11.98}{122}
\emline{126.52}{11.98}{123}{127.09}{10.61}{124}
\emline{127.09}{10.61}{125}{127.59}{8.97}{126}
\emline{127.59}{8.97}{127}{128.34}{5.33}{128}
\emline{128.34}{5.33}{129}{129.73}{7.46}{130}
\emline{129.73}{7.46}{131}{131.07}{9.32}{132}
\emline{131.07}{9.32}{133}{132.34}{10.90}{134}
\emline{132.34}{10.90}{135}{133.54}{12.22}{136}
\emline{133.54}{12.22}{137}{134.68}{13.27}{138}
\emline{134.68}{13.27}{139}{135.76}{14.05}{140}
\emline{135.76}{14.05}{141}{136.77}{14.57}{142}
\emline{136.77}{14.57}{143}{137.72}{14.81}{144}
\emline{137.72}{14.81}{145}{138.61}{14.78}{146}
\emline{138.61}{14.78}{147}{139.43}{14.49}{148}
\emline{139.43}{14.49}{149}{140.19}{13.92}{150}
\emline{140.19}{13.92}{151}{140.89}{13.09}{152}
\emline{140.89}{13.09}{153}{141.52}{11.98}{154}
\emline{141.52}{11.98}{155}{142.09}{10.61}{156}
\emline{142.09}{10.61}{157}{142.59}{8.97}{158}
\emline{142.59}{8.97}{159}{143.34}{5.33}{160}
\put(32.34,5.33){\vector(1,0){131.00}}
\put(41.67,10.33){\vector(1,0){26.00}}
\put(67.67,22.33){\vector(-1,0){26.00}}
\put(102.34,22.33){\vector(1,0){34.00}}
\put(54.67,14.33){\makebox(0,0)[cc]{$\exp(ikx)$}}
\put(54.67,26.33){\makebox(0,0)[cc]{$R\exp(-ikx)$}}
\put(119.34,26.33){\makebox(0,0)[cc]{$\phi(x)\exp(iqx)$}}
\put(148.34,9.33){\makebox(0,0)[cc]{.}}
\put(153.34,9.33){\makebox(0,0)[cc]{.}}
\put(158.34,9.33){\makebox(0,0)[cc]{.}}
\put(163.34,9.33){\makebox(0,0)[cc]{.}}
\put(159.34,1.33){\makebox(0,0)[cc]{$x$}}
\put(76.67,9.33){\makebox(0,0)[cc]{$r,t$}}
\emline{83.67}{22.33}{161}{83.67}{5.33}{162}
\put(83.67,18.67){\vector(1,0){0.2}}
\emline{75.67}{18.67}{163}{83.67}{18.67}{164}
\put(78.34,23.00){\makebox(0,0)[cc]{$X$}}
\end{picture}
\caption{Scattering of a plane wave from a semiinfinite periodic potential.} \label{pers}
\end{figure}
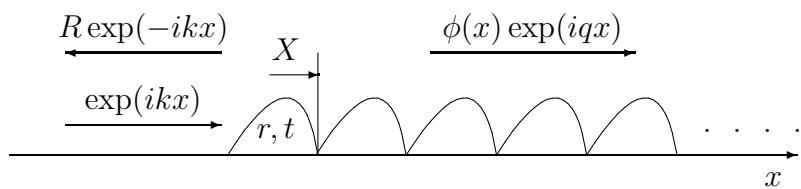

Now we consider one dimensional periodic potentials. Our goal is to find reflection $R_N$
and transmission $T_N$ amplitudes of an
arbitrary periodic potential, containing $N$ periods,
if we know respective amplitudes $r$ and $t$ for a single period.
However we start with semiinfinite periodic potential shown in fig.
\ref{pers}. In fig. \ref{pers} we see an incident wave from the left, and the
reflected wave with an amplitude $R$. We also see the amplitude $X$ of the wave incident
on the second period. For $X$ and $R$ we can immediately write the equations
\begin{equation}
X=t+rRX,\qquad R=r+tRX.
\label{pe2}
\end{equation}
For simplicity we assume that the period is symmetric.
In writing of this system we used (\ref{ee2}) and (\ref{ee4}) for $l=0$ and
took into account that reflection from the semiinfinite potential without
one period is the same as from the whole potential.
Our system has a very interesting symmetry: the equations transform to each
other when we exchange $r$ and $t$.

We can resolve the first equation with respect to $X$ and substitute it into the second.
As a result we obtain the equation
\begin{equation}
R=r+t^2R/(1-rR),
\label{pe3}
\end{equation}
which can be reduced to the form
$y^2-2py+1=0$, where
$y=R$ and $p=(r^2+1-t^2)/2$.
Solution of it is trivial: $y=p-\sqrt{p^2-1}$, however we prefer
to represent it as follows
\begin{equation}
y=\frac{\sqrt{p+1}-\sqrt{p-1}}{\sqrt{p+1}+\sqrt{p-1}},
\label{pe4}
\end{equation}
and we immediately obtain
\begin{equation}
R=\frac{\sqrt{(1+r)^2-t^2}-\sqrt{(1-r)^2-t^2}}{\sqrt{(1+r)^2-t^2}+
\sqrt{(1-r)^2-t^2}}.
\label{pe5}
\end{equation}
We know that a shift by one period inside the periodic potential is accompanied by multiplication
of the wave function by the Bloch factor $\exp(iqa)$. The amplitude $X$ represents the incident
amplitude after shifting inside the potential by one period. Therefore $X$ is precisely this Bloch phase factor:
$X = \exp(iqa)$. Because of symmetry of the system (\ref{pe2}) we can immediately find the Bloch
phase factor from (\ref{pe5}), by exchange of $r$ and $t$:
\begin{equation}
X\equiv e^{iqa}=\frac{\sqrt{(1+t)^2-r^2}-\sqrt{(1-t)^2-r^2}}{\sqrt{(1+t)^2-r^2}+
\sqrt{(1-t)^2-r^2}}.
\label{pe6}
\end{equation}

There are many modifications of these formulas. We present here only one of them.
If we represent
$r = |r|\exp(i\chi_r )$, and $t = \pm i|t|\exp(i\chi_r )$, take into account that
$r^2-t^2 =\exp(2i\chi_r )$,  and substitute into (\ref{pe5}) and (\ref{pe6}),
we obtain
\begin{equation}
R=\frac{\sqrt{\cos\chi_r+|r|}-\sqrt{\cos\chi_r-|r|}}{\sqrt{\cos\chi_r+|r|}+
\sqrt{\cos\chi_r-|r|}},\qquad e^{iqa}=
\frac{\sqrt{\sin\chi_r\mp|t|}-\sqrt{\sin\chi_r\pm|t|}}
{\sqrt{\sin\chi_r\mp|t|}+\sqrt{\sin\chi_r\pm|t|}}.
\label{pe7}
\end{equation}
The obtained formulas contain full the information about Bragg refections. All
these refection are total, and the refection amplitude is a unit complex number $R = \exp(i\phi)$,
when $\cos^2\chi_r -|r|^2< 0$. Positions of the
Bragg peaks are determined by the equation $\chi_r = (n + 1/2)\pi$, the Bragg peaks widths are
determined from the equation $|\Delta\chi| < |r|$. At the Bragg
peaks the Bloch wave vector becomes a complex number $qa = \pi n + iq''a$, which means that the wave
exponentially decays inside the medium.

\subsection{The finite number of periods $N$}

To find $R_N$ and $T_N$ it is sufficient to take into account
that a semiinfinite potential with period $a$ is also periodic with period $Na$, and the Bloch phase
factor of such a periodic potential is $X^N =\exp(iqNa)$. With this in mind we rewrite the system
(\ref{pe2}) in the form
\begin{equation}
X^N=T_N+R_NRX^N,\qquad R=R_N+T_NRX^N,
\label{pe2x}
\end{equation}
where we replaced $r$ and $t$ by $R_N$ and $T_N$. The system (\ref{pe2x}) can be resolved with respect to $R_N$
and $T_N$. As a result we obtain
\begin{equation}
R_N=R\frac{1-\exp(2iqNa)}{1-R^2\exp(2iqNa)},\qquad
T_N=\exp(iqNa)\frac{1-R^2}{1-R^2\exp(2iqNa)}.
\label{pe12}
\end{equation}
These formulas are similar to those for rectangular potential (\ref{ste12}).

\subsection{Practical application}

The method demonstrated here can be used for preparation of mirrors with desirable
properties. In particular, to increase critical angle we need to prepare a supermirror,
which consists of several periodic chains with different periods and different number
of periods~\cite{car}. One can also prepare a monochromatic mirror, which reflects particles
of given energy. For that we need only to prepare a periodic system with some special period,
giving Bragg diffraction at some energy. To exclude Bragg reflection of higher order the
single period of the chain must be of a special form, which can be well calculated~\cite{iigp}.

The method is not restricted to scalar particles. It can be used for neutrons with spin in
magnetic fields~\cite{korn3}, for x-rays, and for elastic waves, and even for
investigation of particles diffusion and moderation.

More over it can be used, and we demonstrate it in the next section,
for description of diffraction in single crystals. With it we have the new dynamical
diffraction theory~\cite{ig10}.

\section{Diffraction in a single crystal}

Interaction of neutrons with media containing atoms in fixed positions is described with the help of
multiple wave scattering (MWS) theory. According to it the neutron wave function scattered from
a single atom at point $\rr=\rr_i$ is
\begin{equation}
\psi(\rr)=\exp(i\kk\rr)-\exp(i\kk\rr_i)\frac{b}{|\rr-\rr_i|}\exp(ik|\rr-\rr_i|),
\label{sph0}
\end{equation}
where $\exp(i\kk\rr)$ is the plane wave of the free incident neutron, and $b$ is scattering amplitude,
which, contrary to the common usage, we define with negative sign.

If we have $N$ atoms, the neutron wave function becomes
\begin{equation}
\psi(\rr)=\exp(i\kk\rr)
-\sum_{n=1}^{N}\psi_n\frac{b_n}{|\rr-\rr_n|}\exp(ik|\rr-\rr_n|),
\label{sph5}
\end{equation}
where $b_n$ is the scattering amplitude of the atom at point $\rr_n$, and
$\psi_n$ is the amplitude of the wave enlightening the $n$-th atom. The amplitude $\psi_n$ is
a superposition of all the waves coming to $n$-th atom from all
the others:
\begin{equation}
\psi_n=\exp(i\kk\rr_n)
-\sum_{n'\ne n}^{}\psi_{n'}\frac{b_{n'}}{|\rr_n-\rr_{n'}|}\exp(ik|\rr_n-\rr_{n'}|).
\label{sph6}
\end{equation}
This is the main system of equations of MWS theory. If we solve it, we find $\psi_{\n}$, substitute into
(\ref{sph5}) and find the final waves scattered by our set of atoms.

\subsection{Scattering from a single crystalline plane}

We can solve this system, when atoms are identical ($b_{\n}=b$) and arranged regularly
on a single infinite plane, which we suppose to coincide with coordinate $(x,y)$-plane.
For simplicity we can imagine the elementary cell of atoms to be a square
with side $a$. From symmetry considerations the solution of the
system (\ref{sph6}) is
\begin{equation}
\psi_{\sn}=C\exp(i\kk\rr_{\sn}),
\label{sph6a}
\end{equation}
where $\rr_{\sn}=a\n$, $\n=(n_x,n_y,0)$ and $n_{x,y}$ are integers.
Substitution of (\ref{sph6a}) into (\ref{sph6}) with $b_{\sn}=b$ gives
$C=1/(1+bS)$,
where $S$ is the sum
\begin{equation}
S=\sum_{\n\ne0}^{}\exp(i\kk_0\rr_{\sn})\frac{\exp(ikr_{\sn})}{r_{\sn}},
\label{sph6c}
\end{equation}
which was calculated in~\cite{ig}, and we do not reproduce it here.

If we substitute (\ref{sph6a}) into (\ref{sph5}), we obtain wave function scattered by the crystalline plane
\begin{equation}
\psi(\rr)=\exp(i\kk\rr)
-Cb\sum_{\n}^{}\frac{\exp(i\kk\rr_{\sn})}{|\rr-\rr_{\sn}|}\exp(ik|\rr-\rr_{\sn}|).
\label{sph51}
\end{equation}
We see that $C$ renormalizes $b$ and in the following we omit it. For interpretation we need to transform
(\ref{sph51}), but for that we need some mathematics.

\subsubsection{Some needed mathematics}
First, we need Fourier expansion of the spherical waves:
\begin{equation}
\eta(r)=\frac{i}{2\pi}\int\limits_{}^{}\frac{d^2p_\|}{p_z}
\exp(i\p_\|\rr+ip_z|z|),
\label{sph8f}
\end{equation}
where $p_z=\sqrt{k^2-p_\|^2}$, $z$ axis is selected along normal to the crystalline plane,
and vector $\p_\|$ has components in $(x,y)$-plane.

Second, we need to know the representation for a sum of arbitrary numbers $f(n)$
\begin{equation}
\sum_{m=N_1}^{N_2}f(m)=\sum_{M=-\infty}^{\infty}\int\limits_{N_1}^{N_2}
f(x)\exp(2\pi iMx)dx.
\label{sum}
\end{equation}
After substitution of (\ref{sph8f}) into (\ref{sph51}) we obtain
\begin{equation}
\psi(\rr)=\exp(i\kk\rr)
-\frac{ib}{2\pi}\sum_{\n}^{}\exp(i\kk\n a)\int\frac{dp_\|^2}{p_z}\exp\left(i\p_\|(\rr-a\n)+ip_z|z|\right).
\label{sph52}
\end{equation}
With the help of (\ref{sum}) the sum is transformed to the other one
\begin{equation}
\psi(\rr)=\exp(i\kk\rr)
-\frac{ib}{2\pi}\sum_{\m}^{}\int d^2n\exp(2\pi\m\n)\exp(i\kk\n a)\int\frac{dp_\|^2}{p_z}
\exp\left(i\p_\|(\rr-a\n)+ip_z|z|\right),
\label{sph53}
\end{equation}
where $\m=(m_x,m_y,0)$ and $m_{x,y}$ are integers. It is easy to integrate (\ref{sph53})
over $d^2n$ and $d^2p$, and as a result we obtain
\begin{equation}
\psi(\rr)=e^{i\kk\rr}
-\sum_{\ta}^{}\frac{2\pi iN_2b}{k_{z\sta}}\exp\left(i\kk_{\sta}\rr+ik_{z\sta}|z|\right)\equiv
\Theta(z<0)\left[\psi_0(\rr)+\psi_r(\rr)\right]+\Theta(z>0)\psi_t(\rr),
\label{sph54}
\end{equation}
where $N_2=1/a^2$ is the density of atoms on the plane,
$\ta=(m_x,m_y,0)/a$ is a vector of the reciprocal plane lattice, $\kk_{\sta}=\kk_\|+\ta$,
and $k_{z\sta}=\sqrt{k^2-\kk_{\sta}^2}$. The representation (\ref{sph54}) is well interpretable.
We see that scattering gives diffracted waves. Some of them,
$e^{i\kk_{\sta}\rr_\|-ik_{z\sta}z}$, go back to the space $z<0$, and they constitute
reflected wave function $\psi_r(\rr)$. The others, $\exp\left(i\kk_{\sta}\rr_\|+ik_{z\sta}z\right)$,
go forward to the space $z>0$ and they constitute
transmitted wave function $\psi_t(\rr)$. Here we use $\Theta$-functions which are 1 or 0, when inequality in their
argument are satisfied or not, respectively.

The set of diffracted waves is infinite, but numerable. The functions $\psi_r(\rr)$, $\psi_t(\rr)$
can be represented as infinitely dimensional vectors
$\psi_r(\rr)=\exp(i\hat{\kk}_{r}\rr)\Psi_r$, $\psi_t(\rr)=\exp(i\hat{\kk}\rr)\Psi_t$,
where $\Psi_{r,t}$ are
infinite columns of complex numbers, which can be expanded
over orthogonal basis $|\ta\rangle$: $\Psi_{r,t}=\sum_{\sta}\alpha_{r,t\sta}|\ta\rangle$.
Here $\alpha_{r,t\sta}$ are expansion coefficients or
coordinates of the vectors $\Psi_{r,t}$ in the basis $|\ta\rangle$. A basis vector $|\ta\rangle$ is an
infinite column, which contains unity
at a position corresponding to $m_x$, $m_y$ of the vector $\ta$, and zero
in all other positions.
Numeration of positions in columns $\Psi$ can be arbitrary, but the position with $m_x=m_y=0$ we numerate
by 0. The operator $\hat{\kk}_{r}\rr$ has eigen vectors $|\ta\rangle$
and eigen values $\kk_{\sta}\rr_\|-k_{z\sta}z$. The operator $\hat{\kk}\rr$ has the same
eigen vectors $|\ta\rangle$
and eigen values $\kk_{\sta}\rr_\|+k_{z\sta}z$.

With these notations the state of the incident particle
can be represented as $\exp(\hat{\kk}\rr)\Psi_0$, where $\Psi_0=|\ta=0\rangle$, the
reflected state is $\Psi_r=\hat r_s\Psi_0$, and the transmitted state is $\Psi_t=\hat t_s\Psi_0$,
where $\hat r_s$ and $\hat t_s=I+\hat r_s$ are reflection and transmission matrices of a single crystalline plane
with matrix elements
\begin{equation}\label{sph57}
\langle\ta|\hat r_s|\ta'\rangle=-i\kappa_{\sta},\quad
\langle\ta|\hat t_s|\ta'\rangle=\delta_{\sta,\sta'}-i\kappa_{\sta},\quad \kappa_{\sta}=2\pi N_2b/k_{z\sta}.
\end{equation}
Here $I$ is
an infinitely dimensional unit matrix, and $\delta_{\sta,\sta'}$ is the Kronecker symbol equal to unity
for $\ta=\ta'$, and 0 otherwise.
The total wave function (\ref{sph54}) can now be represented as
\begin{equation}
\psi(\rr)=\Theta(z<0)\left[\exp(i\hat{\kk}\rr)+\exp(i\hat{\kk}_r\rr)\hat r_s\right]\Psi_0+
\Theta(z>0)\exp(i\hat{\kk}\rr)\hat t_s\Psi_0.
\label{sph56}
\end{equation}
We see that diffraction by a single crystal can be reduced to reflection and transmission of a one dimensional
potential for a particle with an infinitely dimensional spin.

\subsection{Sketch of the dynamical diffraction theory}

Let us imagine the single crystal as a one dimensional system of crystalline planes parallel to
the entrance surface and separated by distance $a$. We can consider it as a one dimensional potential
with symmetric period, consisting of a crystalline plane and two empty spaces of width $a/2$ on both
sides of it. Reflection $\hat r$ and transmission $\hat t$ amplitudes of so defined one period are easily
obtained from $\hat r_s$ and $\hat t_s$ of a single plane:
\begin{equation}\label{sph58}
\hat r=\hat E\hat r_s\hat E, \qquad \hat t=\hat E\hat t_s\hat E=\hat E^2+\hat r,
\end{equation}
where $\hat E$ is a diagonal matrix with matrix elements $\langle \ta|\hat E|\ta\rangle=\exp(ik_{z\sta}a/2)$.
This matrix describes propagation of the state $\Psi$ in empty space.

First, we can find diffraction from a semiinfinite single crystal. We denote by $\hat X$ the state
incident upon the second period of the crystal, and put down equations for $\hat X$ and reflection matrix $\hat R$:
\begin{equation}
\hat X=\hat t+\hat r\hat R\hat X,\qquad \hat R=\hat r+\hat t\hat R\hat X,
\label{pe2n}
\end{equation}
which is similar to (\ref{pe2}). We can resolve the first equation with respect to $\hat X$ and
substitute in the second one. As result, we obtain
\begin{equation}\label{pe2n1}
\hat X=(I-\hat r\hat R)^{-1}\hat t,\qquad\hat R=\hat r+\hat t\hat R(I-\hat r\hat R)^{-1}\hat t,
\end{equation}
however, solution of this system is more complicated than of (\ref{pe2}),
because in general matrices $\hat r$, $\hat R$, $\hat X$ and $\hat t$ do not commute.

Suppose we could find the solution. Then we are able to find diffraction on a single crystal of
finite thickness. We define reflection $\hat R_N$ and transmission $\hat T_N$ amplitudes of the crystal with
$N$ periods, and put down the system of equations similar to (\ref{pe2x}):
$\hat X^N=\hat T_N+\hat R_N\hat R\hat X^N$, $\hat R=\hat R_N+\hat T_N\hat R\hat X^N$.
We can resolve it with respect to
$\hat R_N$ and $\hat T_N$, then we obtain
\begin{equation}\label{emb1}
\hat T_N=(I-\hat R^2)\hat X^N[I-\hat R\hat X^N\hat R\hat X^N]^{-1},\qquad
\hat R_N=[\hat R-\hat X^N\hat R\hat X^N][I-\hat R\hat X^N\hat R\hat X^N]^{-1}.
\end{equation}
In the case of Laue diffraction all the intensity goes through the crystal. It means that reflection can
be neglected, therefore $\hat T_N\approx\hat X^N\approx\hat t^N$.
In the case of Bragg diffraction all the intensity is reflected back to the same
space, from where the incident wave arrived, therefore $\hat X^N$ can be neglected and $\hat R_N\approx\hat R$.

The simplest case is Laue diffraction. To find the wave function after the crystal we need to find eigen function
$\Psi_q$ and eigen values $\lambda_q=\exp(iqa)$ of the matrix $\hat t$: $\hat t\Psi_q=\lambda_q\Psi_q$.

\subsection{Some mathematics of matrices}

Our matrix $\hat r$ has a dyadic structure, which means that
matrix elements of $\hat r$ have the form $\langle\ta_i|\hat r|\ta_j\rangle=a_ib_j$, where
$a_i=-ie_i\kappa_i$, $\kappa_i=2\pi bN_2/k_{zi}$, $k_{zi}=\sqrt{k^2-(\kk_\|+\ta_i)^2}$,
$e_i=\exp(ik_{zi}a/2)$, $b_j=e_j$. Matrix elements of $\hat t$ are $\langle\ta_i|\hat t|\ta_j\rangle=
e_i^2\delta_{ij}+a_ib_j$.

Let's find eigen vectors and eigen values of $\hat t$. Eigen vector is $\sum x_j|\ta_j\rangle$.
The equation is
\begin{equation}\label{sph59}
\hat t\sum x_i|\ta_i\rangle=\lambda\sum x_i|\ta_i\rangle\to e^2x_i+a_i\sum\limits_jb_jx_j=\lambda x_i.
\end{equation}
It follows that $x_i=a_iC/(\lambda-e_i^2)$, where
$C=\sum_jb_jx_j=C\sum_jb_ja_j/(\lambda-e_j^2)$,
from which we get the equation for determination of $\lambda$:
\begin{equation}\label{sph60}
\sum\limits_i\frac{-i\kappa_ie_i^2}{\lambda-e_i^2}\equiv
\sum\limits_{\sta}\frac{-2\pi iN_2b\exp(ik_{z\sta}a)}{k_{z\sta}(\lambda-\exp(ik_{z\sta}a))}=1.
\end{equation}

If $n$ exponents $\exp(ik_{z\sta_j}a)$ are nearly equal, eq. (\ref{sph60}) has $n$
solutions $\lambda_j$ close to each other and to
$\exp(ik_{z\sta_j}a)$. They correspond to $n$ orthogonal eigen states
$\Psi_j=\sum_{i\in n}x_{ji}|\ta_i\rangle$, where summation
is limited mainly to $n$ those $\ta_i$, for which $\exp(ik_{z\sta_j}a)$ are nearly the same.

To find transmitted waves we need to expand the incident wave $\Psi_0$ over $\Psi_j$. As a result we obtain
$\Psi_0=\sum_{j\in n}c_j\Psi_j$.

Transmission of every eigen vector $\Psi_j$ is equal to $\exp(iq_jNa)$. Therefore transmitted wave function is
$$\Psi_t=\sum\limits_{j\in n}\exp(iq_jNa)c_j\Psi_j=\sum\limits_{i,j\in n}c_j\exp(iq_jNa)x_{ji}|\ta_i\rangle,$$
from which we see that there are $n$ transmitted waves $|\ta_j\rangle$ with intensity of the
transmitted wave $I_j=\left|\sum\limits_{i\in n}\exp(iq_jiNa)x_{ij}\right|^2$. We obtained the generalization
of pandell\"osung solution for $n$ waves $|\ta_j\rangle$. In the case of $n=2$ we have $k_{z0}=k_{z\sta}$, when
$|\kk_\||\approx|\kk_\|+\ta|$. This is the standard case of Laue diffraction. If $n=1$, i.e. all $\exp(ik_{z\sta}a)$
are different, we have refraction
with $q=\sqrt{k_z^2-u_0}$, where $u_0=4\pi N_0b$, $N_0=N_2/a=1/a^3$.

For Bragg diffraction we need to know $R$. Let's look at the eq. $R=\hat r+\hat t\hat R(I-\hat r\hat R)^{-1}\hat t$ in
(\ref{pe2n1}). Because of the dyadic form of $\hat r$, we can find that $(1-\hat r\hat R)^{-1}=1+C\hat r\hat R$, where
$C=1/(1-c)$, and $c=\sum_{ij}(b_iR_{ij}a_j)$. With this in mind we can rewrite  eq. for $\hat R$ in the form
\begin{equation}\label{pena}
\hat R-\hat E^2\hat R\hat E^2=\frac{(1+\hat E^2\hat R)\hat r(1+\hat R\hat E^2)}{1-\sum a_iR_{ij}b_j},\quad
R_{ij}=\frac{(a_i+e^2_i\sum_kR_{ik}a_k)(b_j+\sum_nb_nR_{nj}e^2_j)}{(1-e_i^2e_j^2)(1-\sum a_iR_{ij}b_j)}.
\end{equation}
Matrix elements have a structure of the type
$R_{ij}=a_ix_ix_jb_j/(1-e_i^2e_j^2)$ with amplitudes $x_i$, satisfying equations
\begin{equation}\label{pena2}
x_i=\frac{1+x_ie_i^2\sum_kx_ka_kb_k/(1-e_k^2e_i^2)}{\sqrt{1-\sum_{jk}a_jb_jx_ja_kb_kx_k/(1-e_k^2e_i^2)}}.
\end{equation}
We see that the most important are those terms $x_ix_j$, for which $e_i^2e_j^2=1$.
Let's look at two particular cases: specular $i=j=0$, and nonspecular $j\ne i=0$ Bragg reflections.

In the specular case, $i=j=0$, we denote $y=x_0^2$, $r=a_0b_0$, solve algebraic equation
\begin{equation}\label{pena3}
y\left(1-yr^2/(1-e_0^4)\right)=\left(1-e_0^2yr/(1-e_0^4)\right)^2,
\end{equation}
and find $R_{00}=ry/(1-e_0^4)$ identical with (\ref{pe5}), where $t=e_0^2+r$.

In the nonspecular case we multiply two eq-s (\ref{pena2}) for $x_0$ and $x_j$,
denote $y=x_0x_j$, $r_{ij}=a_ib_j$, and obtain the equation
\begin{equation}\label{pena4}
y\left(1-yr_{0j}r_{j0}/(1-e_0^2e_j^2)\right)=\left(1-e_0^2yr_{jj}/(1-e_0^2e_j^2)\right)
\left(1-e_j^2yr_{00}/(1-e_0^2e_j^2)\right).
\end{equation}
After solution of this equation and substitution $R_{j0}=yr_{j0}/(1-e_0^2e_j^2)$ we find for the
reflection amplitude from state $|\ta=0\rangle$ to state $|\ta_j\rangle$ the result:
\begin{equation}\label{pena5}
R_{j0}=\sqrt{\frac{k_{z0}}{k_{z\sta_j}}}\frac{\sqrt{(1+\sqrt{r_{0j}r_{j0}})^2-t_{00}t_{jj}}-
\sqrt{(1-\sqrt{r_{0j}r_{j0}})^2-t_{00}t_{jj}}}{\sqrt{(1+\sqrt{r_{0j}r_{j0}})^2-t_{00}t_{jj}}+
\sqrt{(1-\sqrt{r_{0j}r_{j0}})^2-t_{00}t_{jj}}},
\end{equation}
where we substituted $t_{ii}=e_i^2+r_{ii}$, $r_{ij}=-i\kappa_ie_ie_j$, and $\kappa_j=2\pi N_2b/k_{z\sta_j}$.
The probability of such a reflection is
\begin{equation}\label{pena6}
W_{j0}=\frac{k_{z\sta_j}}{k_{z0}}|R_{j0}|^2=\left|\frac{\sqrt{(1+\sqrt{r_{0j}r_{j0}})^2-t_{00}t_{jj}}-
\sqrt{(1-\sqrt{r_{0j}r_{j0}})^2-t_{00}t_{jj}}}{\sqrt{(1+\sqrt{r_{0j}r_{j0}})^2-t_{00}t_{jj}}+
\sqrt{(1-\sqrt{r_{0j}r_{j0}})^2-t_{00}t_{jj}}}\right|^2,
\end{equation}
where the factor $k_{z\sta_j}/k_{z0}$ gives ratio of currents toward the crystal and away from it.

Because of volume limitations we were not able to present the new dynamical diffraction theory in full details. However
we hope that from our sketch readers are able to understand it. For an exercise we propose to them to answer
the question: what will happen if Bragg and Laue conditions are satisfied simultaneously?
\section*{Acknowledgement}
Author is grateful to the Organizing Committee for invitation, and to Dr. E.P.Shabalin for support.

\end{document}